# Artificial Intelligence (AI) and IT identity: Antecedents Identifying with AI Applications
## *Completed Research*

**Rasha Alahmad**
University of Michigan
rashama@umich.edu

**Lionel Robert**
University of Michigan
lprobert@umich.edu

## Abstract

In the age of Artificial Intelligence and automation, machines have taken over many key managerial tasks. Replacing managers with AI systems may have a negative impact on workers' outcomes. It is unclear if workers receive the same benefits from their relationships with AI systems, raising the question: What degree does the relationship between AI systems and workers impact worker outcomes? We draw on IT identity to understand the influence of identification with AI systems on job performance. From this theoretical perspective, we propose a research model and conduct a survey of 97 MTurk workers to test the model. The findings reveal that work role identity and organizational identity are key determinants of identification with AI systems. Furthermore, the findings show that identification with AI systems does increase job performance.

## Introduction

In the age of Artificial Intelligence (AI) and automation, the role of managers is gradually replaced by algorithms. As a result, many organizations are now completely managed by AI technologies (Petrin, 2019; Robert et al., 2020). Many key managerial tasks have been taken over by machines such as assigning work tasks, evaluating workers' performances, and matching workers and customers (Jarrahi et al., 2019). These technologies have the capability to fulfill the entire spectrum of tasks of highly qualified managers (Susskind & Susskind, 2015). All in all, this new AI approach presents both opportunities and challenges within workplace management.

First, one eminent challenge involves the loss of the interpersonal relationship between managers and their workers, which is often based on identification. Identifying with one's manager, the extent to which the manager is included in the worker's sense of self, has been shown to be particularly important in establishing trust and rapport. Studies have demonstrated that workers who strongly identify with their managers achieve higher levels of organizational commitment (Zhu et al., 2013), increased job satisfaction, and enriched job performance (Hobman et al., 2011). Identification also amplifies the impact of managers on worker's creativity (Wang & Rode, 2010). This is because identifying with managers motivates individuals as far as socioemotionally, psychologically, cognitively, and behaviorally to develop emotional rapport with them and with the organization, which further enhances work outcomes (Hobman et al., 2011; Kark et al., 2003).

Therefore, replacing managers with AI systems may have a negative impact on workers' outcomes. It is simply unclear if workers receive the same benefits from their relationships with AI systems. For instance, Kark et al. (2003) argue how the reliance on this worker-manager identification means that, due to the departure of human managers, the result would be an increased sense of loss and disorientation for workers. This begs the question: What degree does the relationship between the AI system and workers impact worker outcomes? This question speaks directly to the ability of AI-powered organizations to effectively manage their workers.

Similarly, IT identity, which Carter and Grover (2015) define as the extent to which an individual views the use of an IT as integral to his or her sense of self, may be critical to helping us understand the impact of identifying with AI systems. Recent theories suggest that increasing pervasiveness and ubiquity of technologies in organizational settings has led individuals to incorporate technologies into the sense of self as IT identities (Carter & Grover, 2015). If this holds true, then IT identity should be vital to enable us to understand the impact of replacing managers with AI systems. Unfortunately, we currently know very little with regards to the impact of identifying with AI systems on workers' outcomes.

To further ascertain the impact of identification with AI systems on workers" outcomes, we draw on both role identity and organizational identity notions. Role identity is a theoretical approach to understanding the self in relation to social roles (Mishra et al., 2012). Organizational identity represents the degree to which someone's membership in his or her organization is self-defining (Mael & Ashforth, 1992). Although role identity and organizational identity are distinct, individuals occupy roles and belong to groups simultaneously, which directly influence individuals' actions and emotions (Stets & Burke, 2000).

Accordingly, our work posits that role identity and organizational identity have impacts on identifications with the AI system. Our motivation for studying role identity and organizational identity is grounded in literature that implicates the two identities as potent predictors of an individual's emotions and perceptions (Stets & Burke, 2000; Swann et al., 2009). The ways in which individuals view themselves as role occupants and organization members act as a sensemaking filter which then shapes their feelings, frames their experiences, and guides their behaviors.

This paper seeks to develop and empirically examine a theoretical model that explains the impact of role identity and organizational identity on IT identity, where the IT system functions as an AI system. To accomplish this, we conducted a survey study of 97 MTurk workers. The results show that work role identity and organizational identity significantly impacted IT identity. Furthermore, IT identity was an important predictor of job performance. Ultimately, this study makes two significant contributions. First, it presents a theoretical framework of the determinants of IT identity. Second, it empirically tests the determinists and consequences of employee identification with the AI system on job performance.

# Theoretical Background

## *Artificial Intelligence*

AI is defined as "the ability of a machine to perform cognitive functions that we associate with human minds, such as perceiving, reasoning, learning, interacting with the environment, problem solving, decision-making, and even demonstrating creativity" (Rai et al., 2019, p. iii). AI initially appeared as an earliest version of expert systems, but today's AI embodies the ability to act and think rationally and operate autonomously in a way similar to humans (Savić, 2019). This type of technology can assume various forms, such as speech assistants (e.g., Alexa, Siri), online labor platforms (e.g., MTurk, Uber), and recommendation systems (e.g., Amazon, Netflix).

The advancement of AI technologies has led to the emergence of a new generation of digital platforms. These platforms harness the power of cloud computing and machine learning to provide new possibilities for individuals "to sell their labor" (Taylor et al., 2017, p. 25) through interactions with the platforms. Globally, statistics show that 70 million workers are expected to register with these platforms (Heeks, 2017). The utilization of these platforms has grown by 21% from 2016 to 2018 (Online Labour Index, 2019). This technology is becoming more integrated into individuals' everyday lives. For example, online labor platforms (e.g., MTurk) substitute the role of managers by deploying AI techniques (Rai et al., 2019). The direct organizational influence on workers has been replaced by AI systems (Rosenblat, 2018). In this work, we endeavor to provide a new insight into the impact of online labor platforms on workers and their subsequent work outcomes.

## *Identity*

Identity is "a way of organizing information about the self" to define who we are (Clayton, 2003, p. 45). Individuals have different identities that relate to diverse roles they perform (e.g., worker, professor), various social groups they affiliate with (e.g., organizations, work groups), and idiosyncratic personal characteristics they possess (e.g., honest, hardworking) (Burke & Stets, 2009). Identity helps individuals to understand themselves by giving distinct meanings to their experiences and providing guidelines for their behaviors (Gecas, 1982). For instance, a female freelancer may have an understanding of what it means to be ethical when she thinks about herself as a moral person, a notion of what it signifies as productive when she reflects about herself as a worker, and a reliable team member when she sees herself in relation to a specific work group (Stets & Serpe, 2016). These meanings allow her to define herself in terms of a moral person identity, work role identity, and work group identity.

Identity typically includes three basics: role identity, social identity (e.g., organizational identity), and person identity (e.g., IT identity) (Burke & Stets, 2009). This type of classification helps to understand workers' behaviors and actions as they are work role occupants and organizational members. We further discuss these identities in the following sections.

### Role identity

Role identity is defined as the self-view attributed to oneself in relation to a social role (Burke & Tully, 1977). The self is multidimensional (Markus & Wurf, 1987); an individual can have various role identities associated with roles portrayed across different social relationships. Individuals can fulfill at least as many role identities as the number of social positions that they occupy (Hogg et al., 1995). For instance, an individual may serve the role as parent, worker, and volunteer, performed separately or simultaneously, and they often alternate between various roles over time.

Roles are also central to comprehending individuals' behaviors within the workplace. Binyamin (2018) found that workers' service performances were higher when they embraced a strong service role identity. When providing quality service is meaningful to the workers, they are more willing to pay a substantial effort to deliver it. Role identity is considered an intrinsic motivation which enables a vast range of individual behaviors. Farmer et al. findings (2003) revealed that workers with creative role identities were more likely to be imaginative, innovative, and capable of devising novel solutions to problems in the workplace. Furthermore, Grube and Piliavin (2000) investigated the impact of role identity on volunteer performance. They found that a volunteer who devoted considerable time to a number of organizations had the propensity to develop a general volunteer role identity. One's perception of a given role as critical to the success of an organization fosters higher self-esteem and increases one's commitment to that role identity (Grube & Piliavin, 2000).

Besides, role identity theory has been utilized to provide a better understanding of workers' outcomes. To illustrate, Farmer et al. (2003) discovered that workers with a strong creative role identity were more sensitive to contextual support, whereas workers whose role identities were low on creativity were insensitive to leadership support. Following this line of work, Zhang and Bartol (2010) asserted that empowering leadership was more likely to have a strong influence on psychological empowerment when workers considered empowerment as an integral part of their worker role identities.

### Organizational Identity

Social identity is the self as derived from group membership (Tajfel et al., 1971). Organizational identity is a form of social identity, whereby the individual views herself or himself as a member of the organization (Bergami & Bagozzi, 2010). Organizational identity reflects the workers' perceptions of the organization's distinctive values and beliefs. It answers important questions, such as "who are we as an organization?" and "how are we different from other organizations?" (Piening et al., 2016).

Identifying with the organization further activates a feeling of similarity between the organization members and promotes a cohesive sense of "us" and "them" (Burke & Stets, 2009). It motivates the worker to behave in a way that is consistent with the organization's overall norms and standards. Viewing the self as an organizational member also triggers depersonalization symptoms. Rather than considering the self as a singular, distinctive entity, individuals view themselves in terms of the prototypical attributes of the organization (Hogg, 2006). Depersonalization does not mean that the person's identity will be lost; instead, individuals will typically switch back and forth between person identity and organizational identity as the situation demands (Burke & Stets, 2009).

## IT Identity

Information technologies (IT) have become more pervasive in recent years. Countless aspects of everyday life are increasingly mediated by interactions with IT. Recently, researchers have found that interaction with technology is an important source of person identity construction (Carter et al., 2013). As workers interact with technology through social roles and groups, the exchanges with the technology become paramount to the workers' sense of self. IT identity is conceptualized as "the extent to which an individual views use of an IT as integral to his or her sense of self" (Carter & Grover, 2015, p. 932).

Likewise, IT identity draws upon material identity (Dittmar, 2011), where identifying with material objects occurs when these objects are incorporated into the self-concept. Material identity conceptualization is informed by symbolic interactionism (Mead, 1913), which theorizes that developing the self-concept stems from others' perspectives. Person identity (i.e., who we are) is linked to the symbolic significance of objects. Material objects serve as an imaginary viewpoint from which we can see who we are (Dittmar, 1992). For instance, a Mercedes car is often viewed as a symbol of wealth and success. On the basis of symbolic interactionism, the person who looks at herself or himself in relation to one's Mercedes perceives herself or herself as a rich and successful person Thus, imaginary meanings about objects become important to how individuals develop, maintain, and understand themselves (Dittmar, 2011).

Technology is a material object where "an individual consciously engages with, as an end-user, to produce, store, and communicate information; that could be accessible to that person across time and space" (Carter & Grover, 2015, p. 932). IT identity is manifested by overlapping conceptual boundaries between the self and technology, which is experienced as a sense of connectedness with technology. According to IT identity, considering a given technology as an important part of the sense of self expands the self through incorporating technology utilities into the self. IT identity is usually reflected by three key dimensions: relatedness, dependence, and emotional energy. Relatedness refers to the blurring of boundaries between the self and a technology, and it is experienced as a sense of connectedness with the technology. Dependence pertains to the degree of reliance on a technology. Lastly, emotional energy relates to emotional attachment to a given technology. It is reflected by an individual's feelings of enthusiasm and energy when thinking of oneself in relation to technology.

# Research Model

We developed a theoretical research model (Figure 1) to answer the following research questions: 1) What is the impact of work role identity and organizational identity on IT identity? 2) what is the impact of IT identity on job performance?

## *Hypotheses development*

We propose that identifying with the work role leads to more strongly identifying with technology. Work role identity represents the degree to which someone relies on his or her work role to determine who he or she is or how one views oneself (Mishra et al., 2012). A worker who considers her job to be important and delivers with a sense of fulfillment and purpose is more likely to identify herself with the objects associated with her job. She is also more likely to be happy when she uses these objects to perform her work tasks. In fact, research suggests that role identity has some impact on an individual's identification with technology. Carter et al. (2017) found that technology supported individuals' role identities by affording them the opportunity to access more information and options, which induced them to develop a greater sense of connectedness with the technology.

The link between work role identity and IT identity can be explained by the degree of investment in identity (McCall & Simmons, 1966). The importance of a certain role identity largely depends on the degree of investment in that identity (Amatea et al., 1986). Each individual has various identities, and these identities are not all equally important in all situations. In the workplace, work role identity is most likely to be salient and committed to across various work-related contexts. Researchers (e.g., Callero, 1985; Stryker & Serpe, 1982) argued that the higher the commitment to a specific role, the larger the amount of time devoted to that role. Technology, even with the narrow scope of application, can become a critical part of the sense of self if the individual devotes a substantial amount of time to use that technology (Carter & Grover, 2015). Accordingly, the person who considers an AI system as instrumental to his or her work role is more likely to exert more time and effort on AI system use to accomplish one's work role tasks. Thus, the worker identifies with the AI system.

Similarly, the correlation between work role identity and IT identity is aligned with technology dependence. The worker who perceives that past work role tasks were associated with technology use is more likely to connect achieving future role-relevant tasks to that technology. For instance, the Uber driver who perceives that the Uber app makes one's life better and gives more control over one's schedule is more likely to provide Uber rides in the future. Considering technology critical to role task performance increases the degree of dependence on that particular technology. This relationship has been confirmed by Vincent (2006), who found that one's technological involvement induces a user to rely on it to conduct future activities.

Technology activates memories associated with prior work-related interactions which further increases dependence on technology. These memories help workers to better understand themselves and perform their roles (Treem & Leonardi, 2013). Technology can help a person determine the degree of association with others, availability, and visibility. Mazmanian et al. (2006) demonstrated that workers maintained a certain level of availability to sustain their relationships with others. Palen et al. (2000) found that cell phones helped individuals to sustain a certain level of accessibility across roles. Due to the possibility of keeping memories and tracking various social and personal events, workers relied more on technologies when interacting with others (Vincent, 2006).

Carter and Grover (2015) proposed that emotional benefits resulting from past use of technology have a potential to increase the strength of a person's IT identity. This is echoed in the literature where researchers demonstrated that higher levels of satisfaction built greater dependence on the technology (Rai et al., 2002). Furthermore, the sense of enjoyment and satisfaction is activated by these prior experiences with technology, which also increases one's ongoing commitment to using that technology (Bhattacherjee & Premkumar, 2004). As a result, we hypothesize that:

**H1:** Job role identity positively impacts IT identity.

We propose that identifying with organizations leads to identifying with technology in the workplace. Organizational identity represents the degree to which someone's membership in his or her organization is self-defining. By identifying the self with the organization, the person experiences a sense of emotional attachment toward it (Bergami & Bagozzi, 2010). The attachment and belongingness an individual feels should reinforce the pleasure associated with using the organizational objects.

Individuals who identify themselves with the organization are more likely to exert extra efforts (O'Reilly & Chatman, 1986). Studies exhibited that identification with groups in the workplace motivated workers to use technological features more in a repetitive way (Pan et al., 2017). A steady stream of research has shown that organizational image impacts customers' judgments and motivates them to respond in a positive manner (Wansink et al., 1998). Overall, having a positive view towards the organization has been found to yield positive product evaluations (Brown & Dacin, 1997). Indeed, Ahearne et al. (2005) found that the stronger the customers' identification with the organization, the greater the customers' product utilization.

On the basis of previous studies, we suggest that identification with the organization is the key underlying variable which induces workers to identify themselves with the AI system. Workers who identify themselves with organization are more likely to rely on technology and have a greater sense of connection and emotion toward it. Thus, we hypothesize that:

**H2:** Organizational identity positively impacts IT identity.

We propose that identifying with technology at the workplace impacts job performance. Identification is a vital construct in explaining individuals' behaviors (Suh et al., 2011). Studies have demonstrated that identification with objects impacts the intention to use them through emotional attachment (Thomson et al., 2005). Emotional attachment to technology motivates individuals to be more engaged with that technology and enjoy their interactions accordingly (Li et al., 2006). Indeed, You and Robert (2018) found that emotional attachment to robots enhanced team performance and viability. Sykes and Venkatesh (2017) argued that the more IT functionalities the workers accessed, the more likely they become efficient and effective in performing their work tasks. Thus, we hypothesize that:

**H3:** IT identity positively impacts job performance.

# Method

## *The settings and participants*

To test these hypotheses, we conducted a survey on Amazon Mechanical Turk (MTurk). We had a total of 97 MTurk worker survey responses after removing 3 respondents who had incomplete responses, or who completed the survey in less than 3 mins or more than 30 mins which signaled a low level of attention. Forty-three respondents were female, forty-one identified as white, and thirty-seven were married. The average length of MTurk experience was 3 years. The respondents' ages ranged from 21 to 65, with a mean age of 29. The MTurk population tends to be younger than the overall population; 20% of MTurkers were born after 1990, and 60% were born after 1980 (Difallah, Filatova, & Ipeirotis, 2018). The survey was launched on April 25, 2019 and all responses were completed in the same day. Each MTurk worker got $1.00 in exchange for their participation.

## *Measurements*

We used a 3-item 7-point Likert scale adapted from Callero (1985) and Farmer et al. (2003) to measure role identity. Organizational identity was measured using three items developed by Smidts et al. (2001) assessed on seven-point disagree/agree scale. IT identity was evaluated using six items developed by Carter (2013). The scale of job performance consisted of three items adapted from Moqbel et al. (2013) using a 7-point Likert scale. We used several control variables to rule out any potential alternative explanations for the results. We examined gender, age, race, education, and marital status; none of these variables showed statistically significant mean difference.

# Analysis and Results

## *Measurement validity*

We employed SmartPLS 3.2.8, which is a partial least squares (PLS) tool and uses a component-based approach to maximize the estimation of variance in the research model. The reliability of the constructs can be examined through composite reliability (CR), Cronbach's alpha (CA), and average variance extracted (AVE). These values should be greater than 0.7, 0.7 and 0.5 respectively. As shown in Table 1, CR values are acceptable since all of them are above 0.80. AVE ranges from 0.52 to 0.73. These results provide evidence of convergent validity. In addition, we used both the Fornell-Larcker criterion (1981) and the heterotrait-monotrait (HTMT) ratio of correlations (Henseler et al., 2015) to validate the discriminant validity of our model. First, discriminant validity can be verified when the square root of AVE for each construct is greater than its correlation with other constructs. Table 1 depicts that the square root of AVE for each construct is higher than its correlations with all other constructs. Second, Henseler et al. (2015) proposed an alternative, more sensitive method in detecting discriminant validity called the HTMT ratio of correlations. We found that the HTMT values were below the conservative threshold of 0.85. This is further evidence of discriminant validity among the constructs.

| Constructs | Mean | Std. Dev. | CR | CA | 1 | 2 | 3 | 4 |
|---|---|---|---|---|---|---|---|---|
| IT identity | 5.29 | 0.91 | 0.87 | 0.82 | **0.72** | | | |
| Role identity | 5.09 | 1.12 | 0.84 | 0.71 | 0.55* | **0.79** | | |
| Organizational identity | 5.04 | 1.17 | 0.89 | 0.81 | 0.68** | 0.55** | **0.85** | |
| Job performance | 5.13 | 1.20 | 0.87 | 0.79 | 0.64** | 0.57** | 0.60** | **0.84** |
| *Note*: n=97. CA= Cronbach's alpha. Values on the diagonals represent the square root of the AVE for each construct. Values off-diagonal are correlations between constructs. | | | | | | | | |

**Table 1. Descriptive Statistics and Correlations Among Constructs**

## Hypothesis Testing

Our research model was evaluated using PLS. The significance of the path coefficients was examined by employing a bootstrapping with 5,000 iterations. The results in Figure 1 show that the structural model explained 53% of variance in IT identity as well as 44% of variance in job performance. Regarding the impact of work role identity on IT identity, there was a significant positive effect (β= 0.26, t=2.52, p< .05). Furthermore, the path coefficient from organizational identity and IT identity was significant (β= 0.55, t=5.93, p <.05). We also discerned that IT identity significantly impacted job performance (β= 0.67, t=12.6, p<.05).

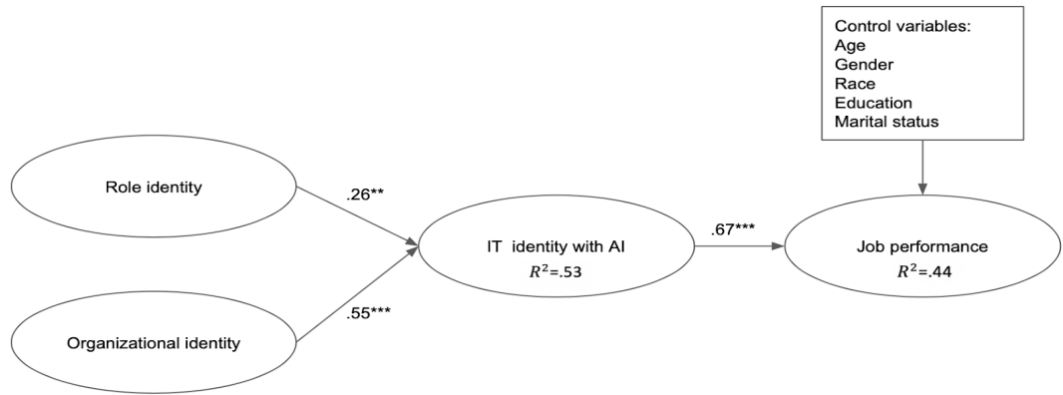

**Figure 1: Results of PLS Structural Model**

## Discussion and Conclusion

The purpose of this research was to present and test a model of antecedents identifying with AI systems. Our findings indicate that identifying with the work role and organization can increase identifying with the labor platform. In addition, we find that identification with the labor platform enhances workers' job performance. Our results are a first step toward demonstrating the importance of identification with AI systems on work outcomes.

The present work examines the role of IT identity as an explanation for how individuals interact with AI systems, answering the call of Vodanovich, Sundaram, and Myers (2010). The study introduces two determinants of IT identity: role identity and organizational identity. We find that both identities explain significant variance in the labor platform identity. Furthermore, our study provides a theoretically informed explanation for the impact of identification with the platform on workers' job performances. To compensate for the absence of human managers in the platform, our study shows that workers develop an identity with the platform itself, and that identity fosters their job performance. The present paper opens a new avenue for future research to assess the impact of identification with AI systems on other work outcomes, such as wellbeing and satisfaction. Future research can empirically test our conceptual model on other labor platforms.

However, this study has several limitations. First, it is a cross-sectional study.; therefore, we cannot infer causal relationships. Longitudinal follow-up may offer additional information, while a comparison of identification with AI systems in two different points of time would help to find more accurate results. Second, there is a large variation in the age of participants. Prior work which compared the MTurk population with the general population found that MTurk workers were younger (Chandler & Shapiro, 2016). There, future work should collect data from other platforms (e.g., Upwork, Uber) to test the model in different work environments and potentially different populations.